\documentclass{cs23proc}

\usepackage{amsmath,gensymb}

\editors{Takeru Suzuki and the Cool Stars 23 Organizing Team}
\publisher{Zenodo}
\conference{The 23th Cambridge Workshop on Cool Stars, Stellar Systems, and the Sun (Cool Stars 23)}
\conferencedate{2026}

\title{Asteroseismic imprints of strong non-axisymmetric fields in the cores of red giants}
\author{Nicholas Z. Rui,$^{1,2,3}$
J. M. Joel Ong,$^4$
Armand Leclerc,$^5$
Daniel Lecoanet,$^{2,6}$
Lisa Bugnet,$^5$
Janosz W. Dewberry,$^7$
Bastien Liagre,$^8$
and St\'ephane Mathis$^9$}

\affiliation{$^{1}$Department of Astrophysical Sciences, Princeton University, 4 Ivy Lane, Princeton, NJ 08544, USA\\
$^{2}$Center for Interdisciplinary Exploration and Research in Astrophysics (CIERA), Northwestern University, 1800 Sherman Ave., Evanston, IL 60201, USA\\
$^{3}$NASA Hubble Fellow\\
$^{4}$Sydney Institute for Astronomy, University of Sydney, A28 Physics Road, Sydney NSW 2006, Australia\\
$^{5}$Institute of Science and Technology Austria, Am Campus 1, Klosterneuburg, 3400, Austria\\
$^{6}$Department of Engineering Sciences and Applied Mathematics, McCormick School of Engineering, Northwestern University, 2145 Sheridan Road, Evanston, IL 60208, USA\\
$^{7}$Department of Astronomy, University of Massachusetts Amherst, 710 N Pleasant St, Amherst, MA 01003, USA\\
$^{8}$Universit\'e Paris Cit\'e, Universit\'e Paris-Saclay, CEA, CNRS, AIM, F-91191 Gif-sur-Yvette, France\\
$^{9}$Universit\'e Paris-Saclay, Universit\'e Paris Cit\'e, CEA, CNRS, AIM, F-91191 Gif-sur-Yvette, France}

\shorttitle{Asteroseismic Magnetometry}
\shortauthors{Rui et al.}

\abs{To date, magnetic fields have been asteroseismically measured in nearly one hundred red giant cores. However, most analyses assume weak magnetic fields and slow rotation so that perturbation theory can be applied. The ``traditional approximation of rotation and magnetism'' (TARM) method can predict gravity-mode frequencies under strong magnetic fields and rapid rotation rates. So far, this formalism requires the magnetic field to be symmetric about the rotation axis.

We generalize the TARM formalism to apply to arbitrary magnetic field geometries, including cases where the magnetic and rotation axes are misaligned, as well as fields with no symmetry axis at all. The resulting gravity modes exhibit a rich diversity of wave behavior, including oblique pulsation and avoided crossings. We also clarify the domains of validity of perturbation theory and the TARM formalism.}

\begin{document}

\maketitle

\section{Introduction}

Asteroseismology has very recently offered the only direct glimpses into magnetic fields in the deep interiors of stars.
The Lorentz forces associated with these magnetic structures increase the frequencies of gravity (g) modes in a distinctive way \citep[e.g.,][]{Mathis:2021:Magnetoasteroseismology,Bugnet:2021:MagneticI} which has to date been used to infer strong near-core magnetic fields in roughly $80$ red giants \citep{Li:2022:30to100kG,Deheuvels:2023:MagneticRG,Li:2023:13Magnetic,Hatt:2024:MagneticRG,Villate:2026:MagneticOffset,Deheuvels:2026:NearCritical} as well as in a handful of main-sequence stars \citep{Vandersnickt:2025:BcepMagnetic,Takata:2026:GDorToroidalField,Ihallaine:2026:MagneticGammaDor}.
Approximately one-fifth of red giants also exhibit suppressed dipole modes \citep{Stello:2016:MagneticPrevalence} which are usually attributed to magnetic suppression \citet{Fuller:2015:SuppressedDipole} by strong magnetic fields.
Magnetic suppression of g modes is expected to occur when the field strength exceeds a frequency-dependent threshold given by \citet{Fuller:2015:SuppressedDipole}:
\begin{equation} \label{eqn:Brcrit}
    B_{r,\mathrm{crit}} \sim \sqrt{\rho}\omega^2r/N\mathrm{,}
\end{equation}
where $\rho$ is the density, $\omega$ is the mode frequency, $r$ is the radial coordinate, and $N$ is the Brunt--V\"ais\"al\"a frequency.
At these field strengths (typically $\approx10^5$--$10^6\,\mathrm{G}$ in red giant cores), magnetic forces become comparable to buoyant ones and are expected to affect efficient dissipation of g-mode energy, although the exact mechanism is not well understood \citep{Loi:2017:AlfvenResonances,Loi:2020:MGPackets,Lecoanet:2017:MagneticConversion,Lecoanet:2022:HD43317,Rui:2023:MagneticSuppression,Mueller:2025:RayTracing,David:2025:Magnetogravity}.

Most asteroseismic observations of magnetically induced frequency shifts are interpreted using perturbation theory, which assumes that magnetic fields are weak \citep[e.g.,][]{Gough:1990:MagneticPertTheory}.
In particular, perturbation theory ignores the ability of strong magnetic fields to significantly deform the structures of g modes themselves, an assumption which fails when the Lorentz force is comparable to buoyancy.
This hypothesis breaks down for stars with ``near-critical'' magnetic fields which are too strong for perturbation theory to be valid but not so strong that magnetic suppression is expected to make oscillation modes unmeasurable.

\citet{Rui:2023:MagneticSuppression} and \citet{Rui:2024:TARM} recently developed a non-perturbative formalism for predicting frequency shifts due to near-critical magnetic fields based on the \textit{traditional approximation of rotation and magnetism} (TARM).
This formalism takes advantage of the short radial wavelengths $\lambda_r=2\pi/k_r$ of g modes in red giants and many main-sequence stars, which follows from the gravity-wave dispersion relation $k_r\approx\sqrt{\ell(\ell+1)}N/\omega r$ where $N/\omega\gg1$ ($\simeq100$ for typically excited modes in red giant cores).
This property allows the perturbed magnetohydrodynamic equations to be approximately separated in the radial and horizontal directions.
This is akin to the standard mathematical treatment for waveguides \citep[e.g.,][]{Jackson:1999:ClassicalElectrodynamics}.
It is also directly analogous to the calculation of g-mode frequencies affected by strong Coriolis forces under the \textit{traditional approximation of rotation} \citep[TAR; e.g.,][]{Hough:1897:HoughFunctions,Hough:1898:HoughFunctions,Bildsten:1996:OceanGModes,Lee:1997:TAR}.
A prerequisite for applying the TARM formalism is that the magnetic field be axisymmetric about the rotation axis.

\citet{Deheuvels:2026:NearCritical} recently applied the TARM formalism to the fitting of $8$ red giants possessing near-critical core magnetic fields.
The detection of non-perturbative frequency shifts allows some constraint on the radial structure of the field, which \citet{Deheuvels:2026:NearCritical} find to be roughly confined to layers which previously belonged to the main-sequence convective core \citep[suggestive of a dynamo origin for the field;][]{Cantiello:2016:EvolvingMagnetic}.
However, our ability to probe the full diversity of magnetic structures in stellar interiors is inhibited by our inability to predict g-mode frequencies under strong, non-axisymmetric magnetic fields.
This Proceeding summarizes our recent progress in extending the TARM formalism to non-axisymmetric magnetorotational structures, with more details given in a recently submitted manuscript \citep{Rui:2026:magnetogeneral}.

\section{Formalism and Results}

The propagation of magnetogravity waves is governed by the linearized magneto-Boussinesq equations \citep{Proctor:1982:Magnetoconvection,Lecoanet:2017:MagneticConversion}:
\begin{subequations} \label{eqn:magnetoboussinesq}
    \begin{gather}
        \begin{aligned}
        \rho_0&\partial_t^2\vec{\xi} + 2\rho_0\vec{\Omega}\times\partial_t\vec{\xi} \\
        &= -\nabla\left(p' + \frac{1}{4\pi}\vec{B}_0\cdot\vec{B}'\right)
        - \rho_0N^2\xi_r\hat{r}
        + \frac{1}{4\pi}\left(\vec{B}_0\cdot\nabla\right)\vec{B}'
        \end{aligned}
        \\
        \nabla\cdot\vec{\xi} = 0\mathrm{,}
    \end{gather}
\end{subequations}
where $p'$, $\vec{\xi}$, and $\vec{B}'$ are the pressure, fluid displacement, and magnetic field perturbations, $\vec{\Omega}=\Omega\hat{z}$ is the rotation vector (rigid rotation has been assumed), and subscripts ``0'' indicate equilibrium quantities.

The content of the TARM is to approximate both $\vec{\Omega}$ and $\vec{B}_0$ as being purely radial:
\begin{subequations} \label{eqn:TARM}
    \begin{gather}
        \vec{\Omega} \simeq \Omega\cos\theta\hat{r} \\
        \vec{B}_0 \simeq B_{0r}(r)\psi(\theta,\phi;r)\hat{r}\mathrm{,}
    \end{gather}
\end{subequations}
where we hereafter suppress the radial argument in $\psi$.
The TARM is justified for small radial wavelengths $\lambda_r$, as long as the horizontal component of $\vec{B}_0$ is not stronger than the radial component by a large factor $\sim\lambda_h/\lambda_r$.
For standing magnetogravity modes for which perturbations have a sinusoidal time dependence $\propto e^{i\omega t}$, we make the following Jefferys--Wentzel--Kramers--Brillouin (JWKB) ansatz:
\begin{subequations} \label{eqn:jwkb_ansatz}
    \begin{gather}
        p' = \rho_0\omega^2r^2\sum_\alpha A_\alpha(r)\pi_\alpha(\theta,\phi;r)e^{-iS_\alpha(r)/\epsilon} \\
        \vec{\xi}_h = r\sum_\alpha A_\alpha(r)\vec{\zeta}_\alpha(\theta,\phi;r)e^{-iS_\alpha(r)/\epsilon}\mathrm{,}
    \end{gather}
\end{subequations}
where the JWKB actions $S_\alpha$ are related to $k_r$ as $k_r=S_\alpha'/\epsilon$.
The dimensionless perturbations $(\pi_\alpha,\vec{\zeta}_\alpha)$ are defined to individually satisfy Equations \ref{eqn:magnetoboussinesq} to lowest-order in $\epsilon$:
\begin{subequations} \label{eqn:transverse}
    \begin{gather}
        \left(1 - b_\alpha^2\psi^2\right)\vec{\zeta}_\alpha - iq\mu\,\hat{r}\times\vec{\zeta}_\alpha - \bar{\nabla}_h\pi_\alpha = 0 \\
        \lambda_\alpha\pi_\alpha + \bar{\nabla}_h\cdot\vec{\zeta}_\alpha = 0\mathrm{,}
    \end{gather}
\end{subequations}
where $q=2\Omega/\omega$ and $b_\alpha=k_{r,\alpha}v_{Ar}/\omega$.
The index $\alpha$ labels families of solutions to the eigenvalue problem in Equations \ref{eqn:transverse}, hereafter referred to as ``magnetogravity polarizations'' in analogy to waves with transverse degrees of freedom.
We solve Equations \ref{eqn:transverse} as a function of $q$ and $b_\alpha$ using the spectral eigenvalue solver of \textit{Dedalus} \citep{Burns:2020:Dedalus}.
For practical purposes, we reparameterize the dimensionless magnetic field strength in terms of
\begin{equation} \label{eqn:a_param}
    a = \frac{b_\alpha}{\sqrt{\lambda_\alpha}} = \left(\frac{N}{\omega}\right)\left(\frac{v_{Ar}/r}{\omega}\right)\mathrm{.}
\end{equation}
The utility of this choice is described in detail in Section 3.2 of \citet{Rui:2024:TARM}.

\begin{figure}
    \centering
    \includegraphics[width=\linewidth]{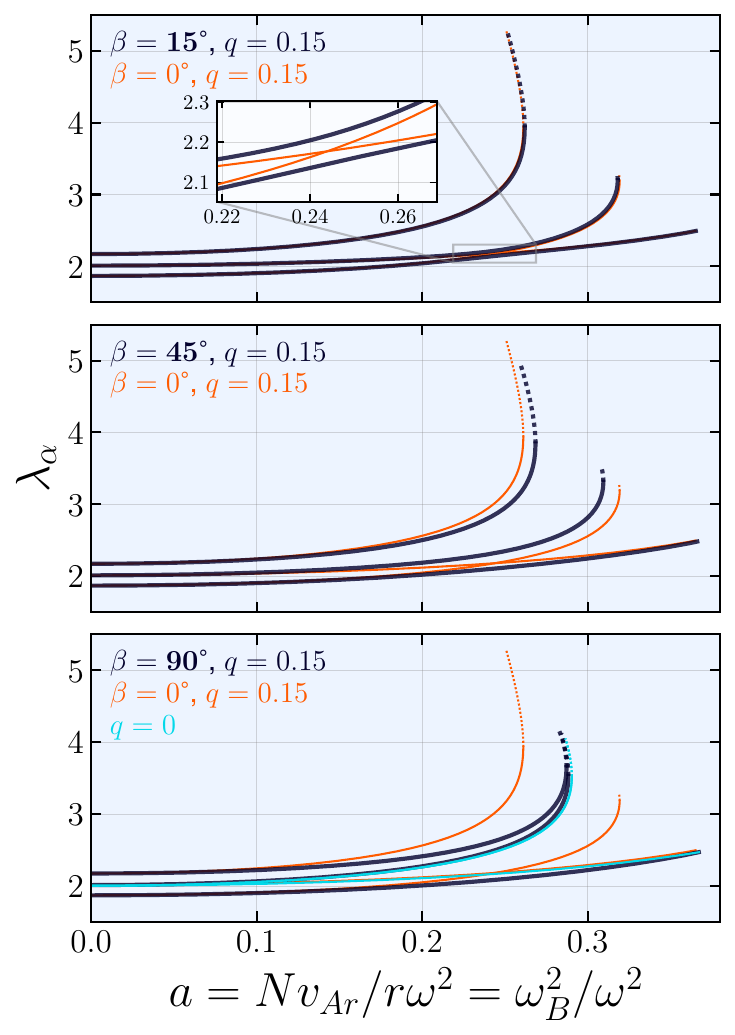}
    \caption{Eigenvalues $\lambda_\alpha$ for dipole magnetogravity polarizations under an inclined dipolar magnetic field (Equation \ref{eqn:dipole_field}).
    Reproduced from Figure 5 of \citet{Rui:2026:magnetogeneral}.}
    \label{fig:FIG_dipole_eigenvalues_1}
\end{figure}

We first consider the simple case of a dipolar magnetic field inclined from the rotation axis by a misalignment angle $\beta$:
\begin{equation} \label{eqn:dipole_field}
    \psi(\theta,\phi) = \sqrt{3}\left(\cos\beta\cos\theta + \sin\beta\sin\theta\cos\phi\right)\mathrm{.}
\end{equation}
Figure \ref{fig:FIG_dipole_eigenvalues_1} shows the eigenvalues $\lambda_\alpha$ for the dipole ($\ell=1$) magnetogravity polarizations for magnetic fields with different values of $\beta$.
For small misalignments, the eigenvalues $\lambda_\alpha$ are very similar to their values in the aligned case ($\beta=0\degree$; orange curves in Figure \ref{fig:FIG_dipole_eigenvalues_1}).
However, unlike in the aligned case, the eigenvalues of the $m=-1$ and $m=0$ magnetogravity polarizations are prevented from crossing by an avoided crossing (shown in the inset in the top panel of Figure \ref{fig:FIG_dipole_eigenvalues_1}).
At this avoided crossing, the two magnetogravity polarizations rapidly exchange character, varying significantly in horizontal structure in the process.
As the misalignment $\beta$ is increased, the avoided crossing broadens until the magnetogravity polarizations are well-separated in $\lambda_\alpha$.

\begin{figure}
    \centering
    \includegraphics[width=\linewidth]{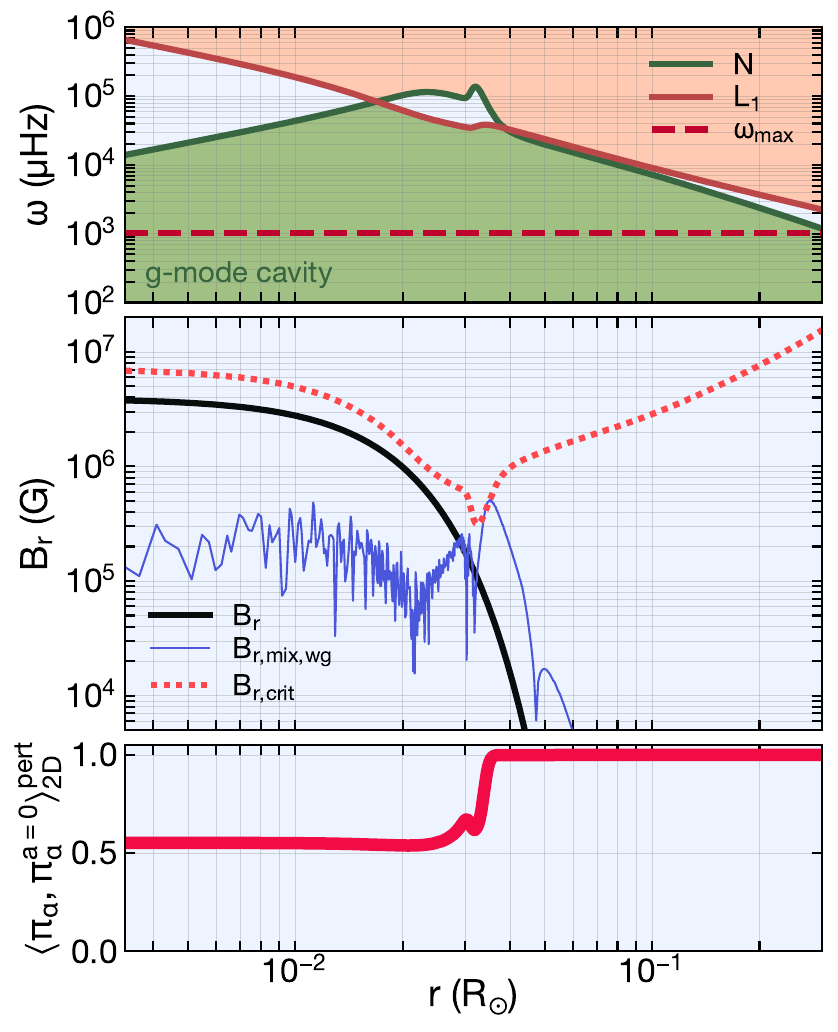}
    \caption{\textit{Top:} Asteroseismic propagation diagram for a $1.2M_\odot$, $5R_\odot$ red giant MESA model.
    \textit{Middle:} $B_{r,\mathrm{mix},\mathrm{wg}}$ (Equation \ref{eqn:jwkb_validity_main}) given a prescribed background magnetic field profile (black), compared to $B_{r,\mathrm{crit}}$ (Equation \ref{eqn:Brcrit}).
    \textit{Bottom:} Assuming an inclined dipolar magnetic field with $\beta=90\degree$, the degree to which a magnetogravity polarization's horizontal structure differs from its structure at $a=0$.
    Reproduced from Figure 11 of \citet{Rui:2026:magnetogeneral}.}
    \label{fig:FIG_Brthresh}
\end{figure}

\citet{Rui:2023:MagneticSuppression} and \citet{Rui:2024:TARM} assumed that magnetogravity waves perfectly follow a single polarization branch as they propagate in radius (the ``adiabatic'' assumption in JWKB theory).
However, retention of higher-order JWKB terms reveals that magnetogravity waves can convert polarizations as they propagate.
Heuristically, ``polarization mixing'' occurs when the eigenvalues $\lambda_\alpha$ are close together and the change in the structure of one polarization geometrically overlaps with the structure of another polarization.
Polarization mixing occurs when the magnitude of the radial component of the magnetic field $B_r$ locally drops below
\begin{equation} \label{eqn:jwkb_validity_main}
    B_{r,\mathrm{mix},\mathrm{wg}} \simeq \frac{4}{[\ell(\ell+1)]^{1/4}}\left(\frac{\omega}{N}\right)^{1/2}\left(\frac{r}{H_{\alpha\beta}}\right)^{1/2}B_{r,\mathrm{crit}}\mathrm{,}
\end{equation}
where $H_{\alpha\beta}$ is the radial scale over which two polarizations $\alpha$ and $\beta$ ``rotate into each other'' as they propagate.
Figure \ref{fig:FIG_Brthresh} shows this quantity for a stellar model of a $1.2M_\odot$, $5R_\odot$ red giant created using Modules for Experiments in Stellar Astrophysics \citep[MESA;][]{Paxton:2011:MESA,Paxton:2013:MESA,Paxton:2015:MESA,Paxton:2018:MESA,Paxton:2019:MESA,Jermyn:2023:MESA}.
Due to the rapid variation in background quantities near the hydrogen burning shell, we find that this polarization mixing is generically expected to occur unless it is geometrically forbidden for symmetry reasons.
The criterion in Equation \ref{eqn:jwkb_validity_main} is related to the breakdown condition for perturbation due to magnetic mixing between modes of different radial orders. 
Near-degeneracy effects associated with this coupling between radial orders will be the subject of a forthcoming study (Liagre et al., in preparation).

\begin{figure}
    \centering
    \includegraphics[width=\linewidth]{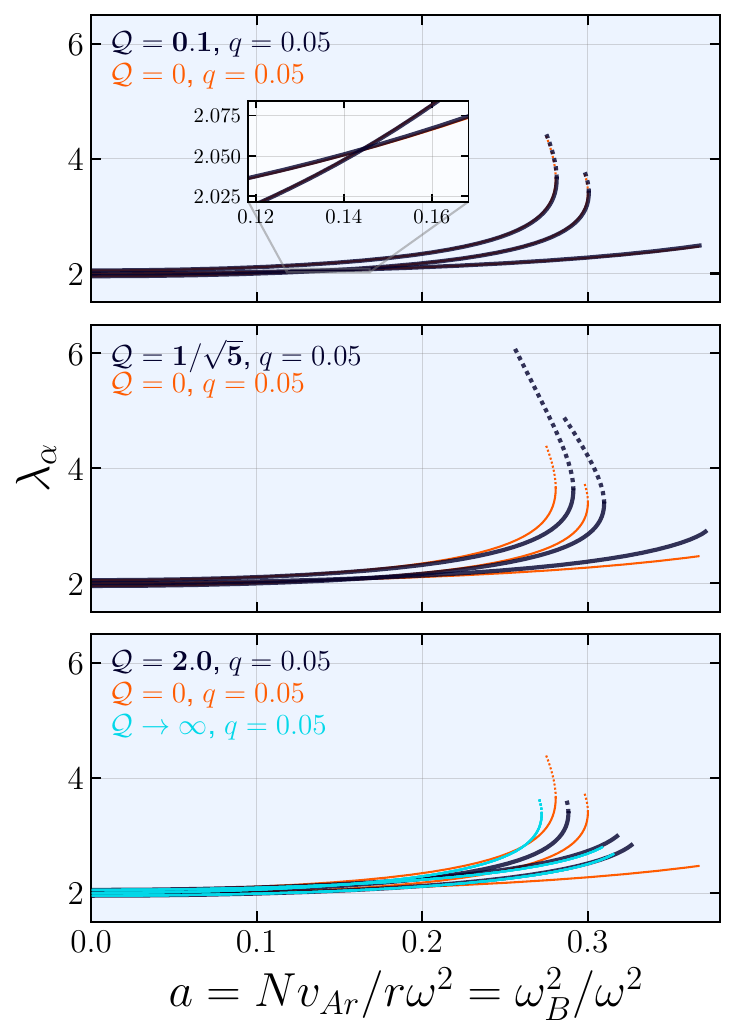}
    \caption{Same as Figure \ref{fig:FIG_dipole_eigenvalues_1} but for a non-axisymmetric dipole-plus-quadrupole geometry (Equation \ref{eqn:dipole_plus_y22}).
    Reproduced from Figure 8 of \citet{Rui:2026:magnetogeneral}.}
    \label{fig:FIG_dpy22_eigenvalues_1}
\end{figure}

\begin{figure}
    \centering
    \includegraphics[width=\linewidth]{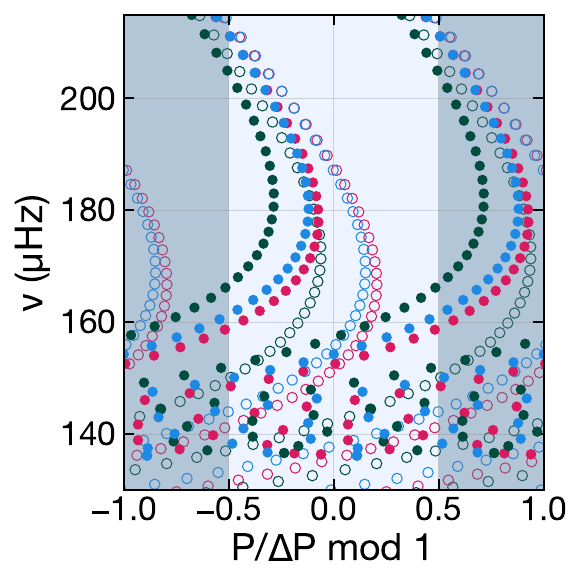}
    \caption{Mock echelle diagram for red-giant dipole g modes which have been strongly perturbed by a non-axisymmetric dipole-plus-quadrupole magnetic field.
    Open circles indicate the predictions of perturbation theory \citep[e.g.,][]{Li:2022:30to100kG}.
    Adapted from Figure 10 of \citet{Rui:2026:magnetogeneral}.}
    \label{fig:PROCFIG_dpy22_echelle}
\end{figure}

However, single-polarization propagation is still approximately expected when symmetry concerns forbid polarization mixing.
The phenomenon of polarization mixing went unnoticed in previous studies \citep[e.g.,][]{Lecoanet:2017:MagneticConversion,Rui:2023:MagneticSuppression,Rui:2024:TARM} precisely because the axisymmetry of the field geometries they studied prevented mixing between polarizations with differing azimuthal quantum number $m$.
However, discrete symmetries can also enforce single-polarization propagation.
We illustratively consider the following field geometry:
\begin{equation} \label{eqn:dipole_plus_y22}
    \psi(\theta,\phi) = \frac{1}{\sqrt{1+\mathcal{Q}^2}}\left(\sqrt{3}\cos\theta + \mathcal{Q}\sqrt{15/4}\sin^2\theta\cos(2\phi)\right)\mathrm{,}
\end{equation}
which is the sum of a real $m=0$ dipole and $m=2$ quadrupole spherical harmonic, with relative weighting given by $\mathcal{Q}$.
Figure \ref{fig:FIG_dpy22_eigenvalues_1} shows the eigenvalues $\lambda_\alpha$ for this non-axisymmetric dipole-plus-quadrupole geometry for different values of $\mathcal{Q}$.
Unlike in the inclined dipole case, the eigenvalues of dipole polarizations under this geometry are allowed to cross.
This is due to the discrete symmetries present in Equation \ref{eqn:dipole_plus_y22}, which prohibit mixing between $\ell=1$ polarizations in spite of the lack of any continuous rotational symmetry.
Under single-polarization propagation, we can straightforwardly calculate predicted mode frequencies by implicitly solving
\begin{equation} \label{eqn:radial_quantization}
    \pi(n+\epsilon_g) = \int_{\mathcal{R}}\frac{\sqrt{\lambda_\alpha(a,q)}N}{\omega r'}\,\mathrm{d}r'
\end{equation}
for $\omega$, where $\mathcal{R}$ denotes the g-mode cavity (note the $\omega$ dependence within $a$ and $q$).
Figure \ref{fig:PROCFIG_dpy22_echelle} presents a mock echelle diagram for our red giant stellar model.
As in \citet{Rui:2024:TARM} for axisymmetric field geometries, we find that perturbation theory underestimates frequency shifts due to near-critical magnetic fields.

\section{Outlook}

The near-critical magnetic red giants discovered by \citet{Deheuvels:2026:NearCritical} are well-fit by axisymmetric magnetic fields embedded deeply within essentially non-rotating cores.
Are these near-critical magnetic red giants representative of a distinct subclass of red giants?
Should we read deeply into their unusually slowly rotating cores, for example in claiming to better understand magnetic angular momentum transport processes \citep{Fuller:2019:SlowingSpins,Skoutnev:2025:MagneticWebs}?
Unfortunately, due to deep selection biases within the current sample of magnetic red giants, the answers to these questions remain elusive.
Theoretical limitations presently prevent us from \textit{discovering} near-critical magnetic red giants whose magnetorotational configurations are outside of a small, especially symmetric subset of all possible geometries.
A more general theory of g-mode asteroseismology under strong magnetic fields will require confronting the complex phenomenology associated with polarization mixing.
This way lies a more complete understanding of the full diversity of magnetorotational structures in stellar interiors.

\section*{Acknowledgments}
N.Z.R. acknowledges support from the NASA Hubble Fellowship grant HST-HF2-51589.001-A awarded by the Space Telescope Science Institute, which is operated by the Association of Universities for Research in Astronomy, Inc., for NASA, under contract NAS5-26555.
J.M.J.O. acknowledges support from the Australian Research Council (ARC) through grants DP210103119 and FL220100117.
L.B. and A.L. gratefully acknowledge support from the European Research Council (ERC) under the Horizon Europe programme (Calcifer; Starting Grant agreement N$^\circ$101165631).
D.L. is partially supported by NSF AAG grant AST-2405812, Sloan Foundation grant FG-2024-21548 and Simons Foundation grant SFI-MPS-T-MPS-00007353.
S.M. acknowledges support from the European Research Council (ERC) under the Horizon Europe programme (Synergy Grant agreement 101071505: 4D-STAR), from the CNES SOHO-GOLF and PLATO grants at CEA-DAp, and from PNPS (CNRS/INSU).
While partially funded by the European Union, views and opinions expressed are, however, those of the authors only and do not necessarily reflect those of the European Union or the European Research Council.
Neither the European Union nor the granting authority can be held responsible for them.

\bibliographystyle{cs23proc}
\bibliography{main.bib}

@ARTICLE{David:2025:Magnetogravity,
       author = {{David}, Cy S. and {Lecoanet}, Daniel and {Garaud}, Pascale},
        title = "{Conversion and Damping of Non-axisymmetric Internal Gravity Waves in Magnetized Stellar Cores}",
      journal = {arXiv e-prints},
         year = 2025,
        month = oct,
          eid = {arXiv:2510.14026},
        pages = {arXiv:2510.14026},
          doi = {10.48550/arXiv.2510.14026},
archivePrefix = {arXiv},
       eprint = {2510.14026},
 primaryClass = {astro-ph.SR},
       adsurl = {https://ui.adsabs.harvard.edu/abs/2025arXiv251014026D}
}

@ARTICLE{Mueller:2025:RayTracing,
       author = {{M{\"u}ller}, Jonas and {Copp{\'e}e}, Quentin and {Hekker}, Saskia},
        title = "{Oscillations of red giant stars with magnetic damping in the core: I. Dissipation of mode energy in dipole-like magnetic fields}",
      journal = {\aap},
         year = 2025,
        month = apr,
       volume = {696},
          eid = {A134},
        pages = {A134},
          doi = {10.1051/0004-6361/202553888},
archivePrefix = {arXiv},
       eprint = {2503.11451},
 primaryClass = {astro-ph.SR},
       adsurl = {https://ui.adsabs.harvard.edu/abs/2025A&A...696A.134M}
}

@ARTICLE{Skoutnev:2025:MagneticWebs,
       author = {{Skoutnev}, Valentin A. and {Beloborodov}, Andrei M.},
        title = "{Magnetic Webs in Stellar Radiative Zones}",
      journal = {\apjl},
         year = 2025,
        month = aug,
       volume = {989},
       number = {1},
          eid = {L4},
        pages = {L4},
          doi = {10.3847/2041-8213/adefda},
archivePrefix = {arXiv},
       eprint = {2504.07223},
 primaryClass = {astro-ph.SR},
       adsurl = {https://ui.adsabs.harvard.edu/abs/2025ApJ...989L...4S}
}

@ARTICLE{Proctor:1982:Magnetoconvection,
       author = {{Proctor}, M.~R.~E. and {Weiss}, N.~O.},
        title = "{REVIEW ARTICLE: Magnetoconvection}",
      journal = {Reports on Progress in Physics},
         year = 1982,
        month = nov,
       volume = {45},
       number = {11},
        pages = {1317-1379},
          doi = {10.1088/0034-4885/45/11/003},
       adsurl = {https://ui.adsabs.harvard.edu/abs/1982RPPh...45.1317P}
}

@ARTICLE{Fuller:2015:SuppressedDipole,
       author = {{Fuller}, Jim and {Cantiello}, Matteo and {Stello}, Dennis and {Garcia}, Rafael A. and {Bildsten}, Lars},
        title = "{Asteroseismology can reveal strong internal magnetic fields in red giant stars}",
      journal = {Science},
         year = 2015,
        month = oct,
       volume = {350},
       number = {6259},
        pages = {423-426},
          doi = {10.1126/science.aac6933},
archivePrefix = {arXiv},
       eprint = {1510.06960},
 primaryClass = {astro-ph.SR},
       adsurl = {https://ui.adsabs.harvard.edu/abs/2015Sci...350..423F}
}

@ARTICLE{Rui:2024:TARM,
       author = {{Rui}, Nicholas Z. and {Ong}, J.~M. Joel and {Mathis}, St{\'e}phane},
        title = "{Asteroseismic g-mode period spacings in strongly magnetic rotating stars}",
      journal = {\mnras},
         year = 2024,
        month = jan,
       volume = {527},
       number = {3},
        pages = {6346-6362},
          doi = {10.1093/mnras/stad3461},
archivePrefix = {arXiv},
       eprint = {2310.19873},
 primaryClass = {astro-ph.SR},
       adsurl = {https://ui.adsabs.harvard.edu/abs/2024MNRAS.527.6346R}
}

@ARTICLE{Rui:2023:MagneticSuppression,
       author = {{Rui}, Nicholas Z. and {Fuller}, Jim},
        title = "{Gravity waves in strong magnetic fields}",
      journal = {\mnras},
         year = 2023,
        month = jul,
       volume = {523},
       number = {1},
        pages = {582-602},
          doi = {10.1093/mnras/stad1424},
archivePrefix = {arXiv},
       eprint = {2303.08147},
 primaryClass = {astro-ph.SR},
       adsurl = {https://ui.adsabs.harvard.edu/abs/2023MNRAS.523..582R}
}

@ARTICLE{Lecoanet:2022:HD43317,
       author = {{Lecoanet}, Daniel and {Bowman}, Dominic M. and {Van Reeth}, Timothy},
        title = "{Asteroseismic inference of the near-core magnetic field strength in the main-sequence B star HD 43317}",
      journal = {\mnras},
         year = 2022,
        month = may,
       volume = {512},
       number = {1},
        pages = {L16-L20},
          doi = {10.1093/mnrasl/slac013},
archivePrefix = {arXiv},
       eprint = {2202.03440},
 primaryClass = {astro-ph.SR},
       adsurl = {https://ui.adsabs.harvard.edu/abs/2022MNRAS.512L..16L}
}

@ARTICLE{Lecoanet:2017:MagneticConversion,
       author = {{Lecoanet}, D. and {Vasil}, G.~M. and {Fuller}, J. and {Cantiello}, M. and {Burns}, K.~J.},
        title = "{Conversion of internal gravity waves into magnetic waves}",
      journal = {\mnras},
         year = 2017,
        month = apr,
       volume = {466},
       number = {2},
        pages = {2181-2193},
          doi = {10.1093/mnras/stw3273},
archivePrefix = {arXiv},
       eprint = {1610.08506},
 primaryClass = {astro-ph.SR},
       adsurl = {https://ui.adsabs.harvard.edu/abs/2017MNRAS.466.2181L}
}

@ARTICLE{Bugnet:2021:MagneticI,
       author = {{Bugnet}, L. and {Prat}, V. and {Mathis}, S. and {Astoul}, A. and {Augustson}, K. and {Garc{\'\i}a}, R.~A. and {Mathur}, S. and {Amard}, L. and {Neiner}, C.},
        title = "{Magnetic signatures on mixed-mode frequencies. I. An axisymmetric fossil field inside the core of red giants}",
      journal = {\aap},
         year = 2021,
        month = jun,
       volume = {650},
          eid = {A53},
        pages = {A53},
          doi = {10.1051/0004-6361/202039159},
archivePrefix = {arXiv},
       eprint = {2102.01216},
 primaryClass = {astro-ph.SR},
       adsurl = {https://ui.adsabs.harvard.edu/abs/2021A&A...650A..53B}
}

@ARTICLE{Cantiello:2016:EvolvingMagnetic,
       author = {{Cantiello}, Matteo and {Fuller}, Jim and {Bildsten}, Lars},
        title = "{Asteroseismic Signatures of Evolving Internal Stellar Magnetic Fields}",
      journal = {\apj},
         year = 2016,
        month = jun,
       volume = {824},
       number = {1},
          eid = {14},
        pages = {14},
          doi = {10.3847/0004-637X/824/1/14},
archivePrefix = {arXiv},
       eprint = {1602.03056},
 primaryClass = {astro-ph.SR},
       adsurl = {https://ui.adsabs.harvard.edu/abs/2016ApJ...824...14C}
}

@ARTICLE{Deheuvels:2023:MagneticRG,
       author = {{Deheuvels}, S. and {Li}, G. and {Ballot}, J. and {Ligni{\`e}res}, F.},
        title = "{Strong magnetic fields detected in the cores of 11 red giant stars using gravity-mode period spacings}",
      journal = {\aap},
         year = 2023,
        month = feb,
       volume = {670},
          eid = {L16},
        pages = {L16},
          doi = {10.1051/0004-6361/202245282},
archivePrefix = {arXiv},
       eprint = {2301.01308},
 primaryClass = {astro-ph.SR},
       adsurl = {https://ui.adsabs.harvard.edu/abs/2023A&A...670L..16D}
}

@ARTICLE{Li:2022:30to100kG,
       author = {{Li}, Gang and {Deheuvels}, S{\'e}bastien and {Ballot}, J{\'e}r{\^o}me and {Ligni{\`e}res}, Fran{\c{c}}ois},
        title = "{Magnetic fields of 30 to 100 kG in the cores of red giant stars}",
      journal = {\nat},
         year = 2022,
        month = oct,
       volume = {610},
       number = {7930},
        pages = {43-46},
          doi = {10.1038/s41586-022-05176-0},
archivePrefix = {arXiv},
       eprint = {2208.09487},
 primaryClass = {astro-ph.SR},
       adsurl = {https://ui.adsabs.harvard.edu/abs/2022Natur.610...43L}
}

@ARTICLE{Takata:2026:GDorToroidalField,
       author = {{Takata}, Masao and {Murphy}, Simon J. and {Kurtz}, Donald W. and {Saio}, Hideyuki and {Shibahashi}, Hiromoto},
        title = "{Asteroseismic detection of a predominantly toroidal magnetic field in the deep interior of the main-sequence F star KIC 9244992}",
      journal = {\mnras},
         year = 2026,
        month = jan,
       volume = {545},
       number = {3},
          eid = {staf2153},
        pages = {staf2153},
          doi = {10.1093/mnras/staf2153},
archivePrefix = {arXiv},
       eprint = {2512.00786},
 primaryClass = {astro-ph.SR},
       adsurl = {https://ui.adsabs.harvard.edu/abs/2026MNRAS.545f2153T}
}

@ARTICLE{Stello:2016:MagneticPrevalence,
       author = {{Stello}, Dennis and {Cantiello}, Matteo and {Fuller}, Jim and {Huber}, Daniel and {Garc{\'\i}a}, Rafael A. and {Bedding}, Timothy R. and {Bildsten}, Lars and {Silva Aguirre}, Victor},
        title = "{A prevalence of dynamo-generated magnetic fields in the cores of intermediate-mass stars}",
      journal = {\nat},
         year = 2016,
        month = jan,
       volume = {529},
       number = {7586},
        pages = {364-367},
          doi = {10.1038/nature16171},
archivePrefix = {arXiv},
       eprint = {1601.00004},
 primaryClass = {astro-ph.SR},
       adsurl = {https://ui.adsabs.harvard.edu/abs/2016Natur.529..364S}
}

@ARTICLE{Hatt:2024:MagneticRG,
       author = {{Hatt}, Emily J. and {Ong}, J.~M. Joel and {Nielsen}, Martin B. and {Chaplin}, William J. and {Davies}, Guy R. and {Deheuvels}, S{\'e}bastien and {Ballot}, J{\'e}r{\^o}me and {Li}, Gang and {Bugnet}, Lisa},
        title = "{Asteroseismic signatures of core magnetism and rotation in hundreds of low-luminosity red giants}",
      journal = {\mnras},
         year = 2024,
        month = oct,
       volume = {534},
       number = {2},
        pages = {1060-1076},
          doi = {10.1093/mnras/stae2053},
archivePrefix = {arXiv},
       eprint = {2409.01157},
 primaryClass = {astro-ph.SR},
       adsurl = {https://ui.adsabs.harvard.edu/abs/2024MNRAS.534.1060H}
}

@ARTICLE{Mathis:2021:Magnetoasteroseismology,
       author = {{Mathis}, S. and {Bugnet}, L. and {Prat}, V. and {Augustson}, K. and {Mathur}, S. and {Garcia}, R.~A.},
        title = "{Probing the internal magnetism of stars using asymptotic magneto-asteroseismology}",
      journal = {\aap},
         year = 2021,
        month = mar,
       volume = {647},
          eid = {A122},
        pages = {A122},
          doi = {10.1051/0004-6361/202039180},
archivePrefix = {arXiv},
       eprint = {2012.11050},
 primaryClass = {astro-ph.SR},
       adsurl = {https://ui.adsabs.harvard.edu/abs/2021A&A...647A.122M}
}

@ARTICLE{Li:2023:13Magnetic,
       author = {{Li}, Gang and {Deheuvels}, S{\'e}bastien and {Li}, Tanda and {Ballot}, J{\'e}r{\^o}me and {Ligni{\`e}res}, Fran{\c{c}}ois},
        title = "{Internal magnetic fields in 13 red giants detected by asteroseismology}",
      journal = {\aap},
         year = 2023,
        month = dec,
       volume = {680},
          eid = {A26},
        pages = {A26},
          doi = {10.1051/0004-6361/202347260},
archivePrefix = {arXiv},
       eprint = {2309.13756},
 primaryClass = {astro-ph.SR},
       adsurl = {https://ui.adsabs.harvard.edu/abs/2023A&A...680A..26L}
}

@ARTICLE{Bildsten:1996:OceanGModes,
       author = {{Bildsten}, Lars and {Ushomirsky}, Greg and {Cutler}, Curt},
        title = "{Ocean g-Modes on Rotating Neutron Stars}",
      journal = {\apj},
         year = 1996,
        month = apr,
       volume = {460},
        pages = {827},
          doi = {10.1086/177012},
       adsurl = {https://ui.adsabs.harvard.edu/abs/1996ApJ...460..827B}
}

@ARTICLE{Lee:1997:TAR,
       author = {{Lee}, Umin and {Saio}, Hideyuki},
        title = "{Low-Frequency Nonradial Oscillations in Rotating Stars. I. Angular Dependence}",
      journal = {\apj},
         year = 1997,
        month = dec,
       volume = {491},
       number = {2},
        pages = {839-845},
          doi = {10.1086/304980},
       adsurl = {https://ui.adsabs.harvard.edu/abs/1997ApJ...491..839L}
}

@ARTICLE{Hough:1897:HoughFunctions,
       author = {{Hough}, S.~S.},
        title = "{On the Application of Harmonic Analysis to the Dynamical Theory of the Tides. Part I. On Laplace's ``Oscillations of the First Species,'' and on the Dynamics of Ocean Currents}",
      journal = {Philosophical Transactions of the Royal Society of London Series A},
         year = 1897,
        month = jan,
       volume = {189},
        pages = {201-257},
          doi = {10.1098/rsta.1897.0009},
       adsurl = {https://ui.adsabs.harvard.edu/abs/1897RSPTA.189..201H}
}

@ARTICLE{Hough:1898:HoughFunctions,
       author = {{Hough}, S.~S.},
        title = "{On the Application of Harmonic Analysis to the Dynamical Theory of the Tides. Part II: On the General Integration of Laplace's Dynamical Equations}",
      journal = {Philosophical Transactions of the Royal Society of London Series A},
         year = 1898,
        month = jan,
       volume = {191},
        pages = {139-185},
          doi = {10.1098/rsta.1898.0005},
       adsurl = {https://ui.adsabs.harvard.edu/abs/1898RSPTA.191..139H}
}

@ARTICLE{Fuller:2019:SlowingSpins,
       author = {{Fuller}, Jim and {Piro}, Anthony L. and {Jermyn}, Adam S.},
        title = "{Slowing the spins of stellar cores}",
      journal = {\mnras},
         year = 2019,
        month = may,
       volume = {485},
       number = {3},
        pages = {3661-3680},
          doi = {10.1093/mnras/stz514},
archivePrefix = {arXiv},
       eprint = {1902.08227},
 primaryClass = {astro-ph.SR},
       adsurl = {https://ui.adsabs.harvard.edu/abs/2019MNRAS.485.3661F}
}

@ARTICLE{Paxton:2011:MESA,
  author = {{Paxton}, B. and {Bildsten}, L. and {Dotter}, A. and {Herwig}, F. and {Lesaffre}, P. and {Timmes}, F.},
  title = {{Modules for Experiments in Stellar Astrophysics (MESA)}},
  journal = {\apjs},
  archivePrefix = {arXiv},
  eprint = {1009.1622},
  primaryClass = {astro-ph.SR},
  year = {2011},
  month = {jan},
  volume = {192},
  eid = {3},
  pages = {3},
  doi = {10.1088/0067-0049/192/1/3},
  adsurl = {https://ui.adsabs.harvard.edu/abs/2011ApJS..192....3P},
}

@ARTICLE{Paxton:2013:MESA,
  author = {{Paxton}, B. and {Cantiello}, M. and {Arras}, P. and {Bildsten}, L. and {Brown}, E.~F. and {Dotter}, A. and {Mankovich}, C. and {Montgomery}, M.~H. and {Stello}, D. and {Timmes}, F.~X. and {Townsend}, R.},
  title = {{Modules for Experiments in Stellar Astrophysics (MESA): Planets, Oscillations, Rotation, and Massive Stars}},
  journal = {\apjs},
  archivePrefix = {arXiv},
  eprint = {1301.0319},
  primaryClass = {astro-ph.SR},
  year = {2013},
  month = {sep},
  volume = {208},
  eid = {4},
  pages = {4},
  doi = {10.1088/0067-0049/208/1/4},
  adsurl = {https://ui.adsabs.harvard.edu/abs/2013ApJS..208....4P},
}

@ARTICLE{Paxton:2015:MESA,
  author = {{Paxton}, B. and {Marchant}, P. and {Schwab}, J. and {Bauer}, E.~B. and {Bildsten}, L. and {Cantiello}, M. and {Dessart}, L. and {Farmer}, R. and {Hu}, H. and {Langer}, N. and {Townsend}, R.~H.~D. and {Townsley}, D.~M. and {Timmes}, F.~X.},
  title = {{Modules for Experiments in Stellar Astrophysics (MESA): Binaries, Pulsations, and Explosions}},
  journal = {\apjs},
  archivePrefix = {arXiv},
  eprint = {1506.03146},
  primaryClass = {astro-ph.SR},
  year = {2015},
  month = {sep},
  volume = {220},
  eid = {15},
  pages = {15},
  doi = {10.1088/0067-0049/220/1/15},
  adsurl = {https://ui.adsabs.harvard.edu/abs/2015ApJS..220...15P},
}

@ARTICLE{Paxton:2018:MESA,
  author = {{Paxton}, B. and {Schwab}, J. and {Bauer}, E.~B. and {Bildsten}, L. and {Blinnikov}, S. and {Duffell}, P. and {Farmer}, R. and {Goldberg}, J.~A. and {Marchant}, P. and {Sorokina}, E. and {Thoul}, A. and {Townsend}, R.~H.~D. and {Timmes}, F.~X.},
  title = {{Modules for Experiments in Stellar Astrophysics (MESA): Convective Boundaries, Element Diffusion, and Massive Star Explosions}},
  journal = {\apjs},
  archivePrefix = {arXiv},
  eprint = {1710.08424},
  primaryClass = {astro-ph.SR},
  year = {2018},
  month = {feb},
  volume = {234},
  eid = {34},
  pages = {34},
  doi = {10.3847/1538-4365/aaa5a8},
  adsurl = {https://ui.adsabs.harvard.edu/abs/2018ApJS..234...34P},
}

@ARTICLE{Paxton:2019:MESA,
       author = {{Paxton}, Bill and {Smolec}, R. and {Schwab}, Josiah and {Gautschy}, A. and
         {Bildsten}, Lars and {Cantiello}, Matteo and {Dotter}, Aaron and
         {Farmer}, R. and {Goldberg}, Jared A. and {Jermyn}, Adam S. and
         {Kanbur}, S.~M. and {Marchant}, Pablo and {Thoul}, Anne and
         {Townsend}, Richard H.~D. and {Wolf}, William M. and {Zhang}, Michael and
         {Timmes}, F.~X.},
        title = "{Modules for Experiments in Stellar Astrophysics (MESA): Pulsating Variable Stars, Rotation, Convective Boundaries, and Energy Conservation}",
      journal = {\apjs},
         year = "2019",
        month = "Jul",
       volume = {243},
       number = {1},
          eid = {10},
        pages = {10},
          doi = {10.3847/1538-4365/ab2241},
archivePrefix = {arXiv},
       eprint = {1903.01426},
 primaryClass = {astro-ph.SR},
       adsurl = {https://ui.adsabs.harvard.edu/abs/2019ApJS..243...10P}
}

@ARTICLE{Jermyn:2023:MESA,
       author = {{Jermyn}, Adam S. and {Bauer}, Evan B. and {Schwab}, Josiah and {Farmer}, R. and {Ball}, Warrick H. and {Bellinger}, Earl P. and {Dotter}, Aaron and {Joyce}, Meridith and {Marchant}, Pablo and {Mombarg}, Joey S.~G. and {Wolf}, William M. and {Sunny Wong}, Tin Long and {Cinquegrana}, Giulia C. and {Farrell}, Eoin and {Smolec}, R. and {Thoul}, Anne and {Cantiello}, Matteo and {Herwig}, Falk and {Toloza}, Odette and {Bildsten}, Lars and {Townsend}, Richard H.~D. and {Timmes}, F.~X.},
        title = "{Modules for Experiments in Stellar Astrophysics (MESA): Time-dependent Convection, Energy Conservation, Automatic Differentiation, and Infrastructure}",
      journal = {\apjs},
         year = 2023,
        month = mar,
       volume = {265},
       number = {1},
          eid = {15},
        pages = {15},
          doi = {10.3847/1538-4365/acae8d},
archivePrefix = {arXiv},
       eprint = {2208.03651},
 primaryClass = {astro-ph.SR},
       adsurl = {https://ui.adsabs.harvard.edu/abs/2023ApJS..265...15J}
}

@ARTICLE{Burns:2020:Dedalus,
       author = {{Burns}, Keaton J. and {Vasil}, Geoffrey M. and {Oishi}, Jeffrey S. and {Lecoanet}, Daniel and {Brown}, Benjamin P.},
        title = "{Dedalus: A flexible framework for numerical simulations with spectral methods}",
      journal = {Physical Review Research},
         year = 2020,
        month = apr,
       volume = {2},
       number = {2},
          eid = {023068},
        pages = {023068},
          doi = {10.1103/PhysRevResearch.2.023068},
archivePrefix = {arXiv},
       eprint = {1905.10388},
 primaryClass = {astro-ph.IM},
       adsurl = {https://ui.adsabs.harvard.edu/abs/2020PhRvR...2b3068B}
}

@ARTICLE{Villate:2026:MagneticOffset,
       author = {{Villate}, Matisse and {Deheuvels}, S{\'e}bastien and {Ballot}, J{\'e}r{\^o}me},
        title = "{Seismic detection of core magnetic fields in red giants using the gravity offset}",
      journal = {arXiv e-prints},
         year = 2026,
        month = feb,
          eid = {arXiv:2602.14570},
        pages = {arXiv:2602.14570},
          doi = {10.48550/arXiv.2602.14570},
archivePrefix = {arXiv},
       eprint = {2602.14570},
 primaryClass = {astro-ph.SR},
       adsurl = {https://ui.adsabs.harvard.edu/abs/2026arXiv260214570V}
}

@ARTICLE{Loi:2020:MGPackets,
       author = {{Loi}, Shyeh Tjing},
        title = "{Magneto-gravity wave packet dynamics in strongly magnetized cores of evolved stars}",
      journal = {\mnras},
         year = 2020,
        month = apr,
       volume = {493},
       number = {4},
        pages = {5726-5742},
          doi = {10.1093/mnras/staa581},
archivePrefix = {arXiv},
       eprint = {2002.11130},
 primaryClass = {astro-ph.SR},
       adsurl = {https://ui.adsabs.harvard.edu/abs/2020MNRAS.493.5726L}
}

@ARTICLE{Loi:2017:AlfvenResonances,
       author = {{Loi}, Shyeh Tjing and {Papaloizou}, John C.~B.},
        title = "{Torsional Alfv{\'e}n resonances as an efficient damping mechanism for non-radial oscillations in red giant stars}",
      journal = {\mnras},
         year = 2017,
        month = may,
       volume = {467},
       number = {3},
        pages = {3212-3225},
          doi = {10.1093/mnras/stx281},
archivePrefix = {arXiv},
       eprint = {1701.08771},
 primaryClass = {astro-ph.SR},
       adsurl = {https://ui.adsabs.harvard.edu/abs/2017MNRAS.467.3212L}
}

\end{document}